\documentclass[aps,prb,twocolumn,showpacs]{revtex4-1}
\usepackage{graphicx}
\usepackage{dcolumn}
\usepackage{bm}
\usepackage{amsmath}
\usepackage{multirow}
\usepackage{tabularx}
\usepackage{rotating}  % used by me
\usepackage{hyperref}  % used by me
\begin{document}
\title{Systematic analysis of structural and magnetic properties of
spinel $CoB_2O_4$(B=Cr,Mn and Fe)compounds from their electronic
structures}  
\author{Debashish Das} %
\affiliation{Department of Physics, Indian Institute of Technology
Guwahati, Guwahati, Assam 781039, India} %

\author{Rajkumar Biswas} %
\affiliation{Department of Physics, Indian Institute of Technology
Guwahati, Guwahati, Assam 781039, India} %

\author{ Subhradip Ghosh}
\affiliation{Department of Physics, Indian Institute of Technology
Guwahati, Guwahati, Assam 781039, India} %

\date{\today}

\begin{abstract}
The structural and magnetic properties of spinel compounds $CoB_2O_4$ (B=Cr,Mn and Fe) 
are studied using the DFT+U 
method and generalized gradient approximation (GGA). We concentrate on understanding
the trends in the properties of these materials as the B cation changes, in terms of 
relative strengths of crystal fields and exchange fields  through an analysis of their electronic
densities of states. We find that the electron-electron correlation plays a significant role in
obtaining the correct structural and electronic ground states. Significant structural distortion in
CoMn$_{2}$O$_{4}$ and "inverted" sublattice occupancy in CoFe$_{2}$O$_{4}$ affects the
magnetic exchange interactions substantially. The trends in the magnetic exchange interactions are analysed in terms of the structural parameters and the 
features in their 
electronic structures. We find that the Fe states in
CoFe$_{2}$O$_{4}$ are extremely localised, irrespective of the symmetry of the
site, which makes it very different from the features of the states of the
B cations in other two compounds. These results provide useful insights into
the trends in the properties of CoB$_{2}$O$_{4}$ compounds with variation
of B cation which would help in understanding the results of recent 
experiments on doping of Mn and Cr in multiferroic CoCr$_{2}$O$_{4}$.
\end{abstract}

\maketitle
\section{Introduction}
Semiconductor oxides in spinel structure have drawn considerable attention
over many decades as apart from their potentials as technologically
important materials \cite{spinel1,spinel2,spinel3,spinel4,spinel5}
they serve as a class of materials to understand 
fundamental physics of magnetic interactions, electron-electron correlations
and coupling of various degrees of freedom. Magnetic spinels with different
magnetic cations at crystallographic inequivalent sites are interesting 
due to the challenges they pose in understanding the complexities that arise
out of different magnetic interactions. The technological importances
of magnetic spinels have been reaffirmed by the recent discovery of 
multiferrocity in CoCr$_{2}$O$_{4}$ \cite{cocr2o4}.Various collinear and
non-collinear structures populate the phase diagram \cite{cocr2o4-1,cocr2o4-2}
of this material as a direct consequence of interplay of magnetic exchanges, 
electron-electron correlation and structural distortions. In order to 
explore possible multi-functionalities in this multiferroic, very
recently, attempts were made to study material properties by adding a 
third magnetic atom. The substitution of Cr by Fe resulted in new phenomena 
like magnetic compensation,sign reversal of Exchange
Bias (EB) effect \cite{padam-apl,padam-aip,padam-thesis} at a critical
temperature and significant magnetostriction\cite{pss13}. Recent structural
and magnetic studies \cite{jap15} upon substitution of Cr by Mn in
CoCr$_{2}$O$_{4}$ too showed a composition and temperature induced 
magnetisation compensation, along with a composition dependent structural
transformation. 

These results imply that the substitution of Cr by another magnetic atom
renormalises the intra and inter-sublattice magnetic interactions and can
be directly connected to the novel magnetism related phenomena. However,
in order to understand the basic physics of such complex phenomena and 
interpret the experimental observations, one need to have a clearer
understanding at the microscopic level. Such understanding can be achieved
if the end compounds that is CoCr$_{2}$O$_{4}$,CoMn$_{2}$O$_{4}$ and
CoFe$_{2}$O$_{4}$ are investigated in a systematic manner and various
physical properties of these materials are understood at a microscopic level,
that is, from their electronic structures.

The standardised parameter-free approach to calculate the electronic structures
and the physical properties is the various implementations of the
Density Functional theory (DFT)\cite{dft}.There is no systematic 
exploration of the compounds under consideration here using DFT based
electronic structure methods. There have been very few DFT based calculations
of CoCr$_{2}$O$_{4}$ \cite{ederer,dasjpd} and of CoFe$_{2}$O$_{4}$ \cite
{ederer1,jphysd,biplab}only. These calculations provided important insights
into the interrelations of sub-lattice occupancies, structural properties
and magnetic properties of these materials on the basis of their 
electronic structures.

In this communication, we present a systematic study of the three compounds
Co$B_{2}$O$_{4}$ (B=Cr,Mn,Fe) with the focus on understanding the 
similarities and differences in their electronic structures, and thus the
structural and magnetic properties in particular. We do these using DFT based
calculations. Detailed information on the structural parameters, their
influences on the electronic structures, the variations in the crystal fields
and exchange fields and their effects on the magnetic properties as one
changes the B element are provided. These results serve as important
prerequisites to analyse the systems when they are doped with a third 
magnetic atom. The paper is organised as follows: section II details the 
computational tools used, section III provides and analyses the results
followed by the conclusions.  
\begin{figure}[ht]
\includegraphics[width=5cm,  keepaspectratio]{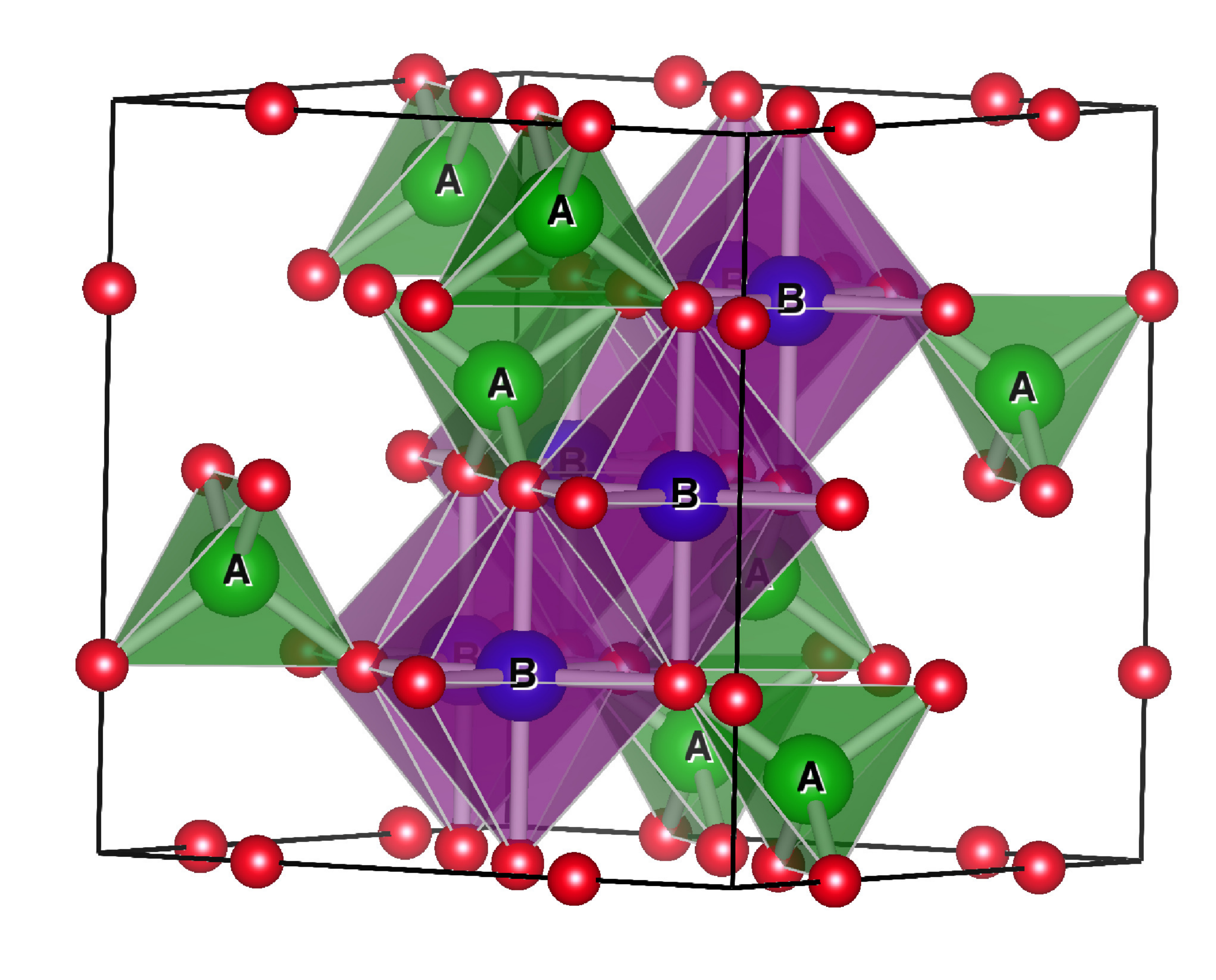}
\caption{\label{Spinel-Figure}  The crystal structure of spinel compounds $AB_{2}O_{4}$.The $A$,$B$ and oxygen 
atoms are shaded in green,purple and red respectively.}
\end{figure}

\section{Computational Details}
We have used DFT+U \cite{dftu} method for all our calculations in order to
address the electron electron strong correlation effects which is 
necessary for oxide systems. Among various different adoptions of DFT+U,
we have used the approach of Dudarev {\it et al} 
\cite{dudarev} where the effects of on-site Coulomb correlation and 
Hund's coupling are represented through an effective
parameter $U_{eff}=U-J$, $U$ being the strength of the Coulomb interaction 
and $J$ the Hund's coupling. In our
calculations, $J$ has been kept at 1 eV as previous DFT calculations 
for a number of transition metal oxides showed that $J$
remains nearly constant for the entire transition metal series \cite{J}. 
The coulomb parameter $U$ has been taken to be 3 eV for Cr, 4 eV 
for Mn and Fe, 5 eV for Co. The rationale behind such choices have been explained
earlier\cite{dasjpd}. The self-consistent quantum mechanical 
equations are solved using the projector augmented wave (PAW) \cite{paw}
basis set as implemented in VASP code \cite{vasp}. 
The exchange correlation part of the Hamiltonian was treated with the
PBE-GGA \cite{pbe} functional. A plane wave cut off of 550 eV and a 
$5 \times 5 \times 5$ mesh centerd at $\Gamma$
point for Brillouin zone integrations have been used throughout to
ensure an energy convergence of $10^{-7}$ eV. For structural relaxations 
the convergence criteria of force on each atom was set to be 
$10^{-4}$ eV/$\AA$. 
 
\section{Results and Discussions}
\subsection{Modeling of the sub-lattice occupancy and spin configuration}
The spinel structure belongs to cubic space group $F d \bar{3}m$ with two
types of cation sublattice-$A$ and $B$. The $A$ sublattice has tetrahedral
symmetry and is usually occupied by a cation in +2 state. The $B$ sublattice
has octahedral symmetry and is usually occupied by a cation in +3 state
(Figure 1). One formula unit of spinel consists of one $A$ sublattice and two
$B$ sublattices. In case of one $B$ atom occupying the tetrahedral 
positions by displacing the $A$ atom to the octahedral sites, the structure
is known as 'Inverse Spinel'. Among the three compounds under consideration,
the CoCr$_{2}$O$_{4}$ is found to be crystallising in
'Normal spinel' structures \cite{menyuk}. In case of CoMn$_{2}$O$_{4}$,
although the sub-lattice occupancy conforms to that of spinel, the crystal
structure is tetragonal with space group $I4_{1}/amd$ \cite{CoMn2O4-mag}. 
CoFe$_{2}$O$_{4}$,on the other hand,has no structural distortion,but
has the sub-lattice occupancy alike Inverse spinel \cite {CoFe2O4-mag}.The
sub-lattice occupancies in CoFe$_{2}$O$_{4}$
is found out to be crucial in determining whether the ground state is insulator
or half-metal \cite{ederer1,jphysd}. Regarding the ground state magnetic
structure, Neutron diffraction studies have shown that CoCr$_{2}$O$_{4}$
and CoMn$_{2}$O$_{4}$ have canted spin structures \cite{CoCr2O4-mag,
CoMn2O4-mag}while CoFe$_{2}$O$_{4}$ has a ferrimagnetic collinear structure,
with the moments at $A$ and $B$ sites anti-aligning, giving rise to the
Neel configuration \cite{CoFe2O4-mag}.Mossbauer studies \cite{CoFe2O4-moss}
later confirmed this.  

The sub-lattice occupancy for CoCr$_{2}$O$_{4}$ and CoMn$_{2}$O$_{4}$ have
been taken to be like normal spinels as have been found from 
experiments. For CoFe$_{2}$O$_{4}$, inverse spinel sub-lattice occupancy 
with maximum inversion has been considered. In absence of disorder at
a particular sub-lattice, this configuration is found to be having the
lowest energy\cite{jphysd}. Since modeling of the sub-lattice disorder
requires either construction of a supercell in the present approach or
consideration of a mean field configuration averaging procedure
\cite{biplab}, we have considered the complete inverse spinel type
sub-lattice occupancy. 
Since the motivation behind this work is a systematic understanding of
the three systems as the chemical identity of the $B$ atom changes, we
have taken the magnetic configurations of the three systems to be collinear.
Although CoCr$_{2}$O$_{4}$ and CoMn$_{2}$O$_{4}$ have non-collinear 
magnetic ground states, the investigations into the collinear magnetic 
states would provide important qualitative insights. In order to find out
the relative alignments of the spins at various sub-lattices, we have taken
2 formula units of the unit cell and done total energy calculations on
different collinear spin configurations. We found the Neel configuration to
be energetically lowest in case of CoCr$_{2}$O$_{4}$ and CoFe$_{2}$O$_{4}$.
In case of CoMn$_{2}$O$_{4}$, we found a magnetic configuration in which
the spins at the sites of a given sub-lattice are anti-parallel, giving rise
to a zero magnetic moment at each sub-lattice, lowest in energy. However,
the energy of the Neel configuration is higher by about only 1 meV per atom.
This result indicates that the actual magnetic ground state of 
CoMn$_{2}$O$_{4}$ would be
frustrated. In subsequent analysis we have considered only the Neel 
configuration therefore.

\subsection{Structural parameters}  %Section - 1.3 
The lattice parameters obtained after full structural relaxations using
DFT+U for the three compounds are given 
in Table \ref{struc-tab}. Figure 2 provides the details of various bond
lengths and bond angles, thus providing a close-up view of the structures
around the tetrahedral and the octahedral sites. The lattice parameters
agree well with the available experimental results. CoCr$_{2}$O$_{4}$
stabilises in the cubic structure with almost no local distortion at the
tetrahedral site as can be seen from the ideal spinel value of $109.47^{0}$
for the $O-Co-O$ angle. There is slight local distortion at the octahedral
site; the $O-Cr-O$ bond angle is $84.55^{0}$, deviating from the ideal value
of $90^{0}$ and the $Co-O-Cr$ bond angle is $121.53^{0}$ in place of ideal
value of $125^{0}$. CoMn$_{2}$O$_{4}$, on the other hand, has a tetragonal
crystal structure with $c/a=1.16$. Subsequently various bond distances
and bond angles (FIG. 1(b)) are significantly dispersed. One can now see
substantial distortion associated with the tetrahedral sites as the $O-Co-O$
bond angles are $108.14^{0}\pm 4.13^{0}$. The octahedra too distorts
considerably with elongation(contraction) of $Mn-O$ bonds along $z(xy)$
directions[FIG. 3(b)]. This leads to significant variations in the 
$O-Mn-O$ and $Mn-O-Co$ angles. The differences in the distortions in the
octahedra around the B sites in CoMn$_{2}$O$_{4}$ in comparison to CoCr$_{2}
$O$_{4}$ is pictorially represented in FIG. 3. In case of CoFe$_{2}$O$_{4}$,
our calculations produce a slight departure from the perfect cubic structure,
the $c-$axis elongated by about $0.5 \%$. This slight loss of cubic symmetry
is due to different distortions of the octahedra around Co and Fe sites.
FIG. 3(c) shows that while there is slight elongations of $Co-O$ bonds along
$z$-direction, the $Fe-O$ bonds in the same direction contracts in comparison
to the same bonds in the $xy$ planes. This gives rise to the small 
tetragonality. The bond angles (FIG. 2(c)) associated with the octahedral and
tetrahedral sites clearly show that the distortions associated with Co atoms
in the octahedral sites. The $O-Co-O$ bond angle deviates to $88.23^{0}\pm
2^{0}$ as opposed to $O-Fe-O$ bond angle of $90.24^{0}\pm0.07^{0}$ which is
quite close to ideal spinel value of $90^{0}$. Similarly, $Fe_{O}-O-Fe_{T}$
bond angle is $124.16^{0}$, while that of $Co_{O}-O-Fe_{T}$ bond angle is
$121.53^{0}$, a larger deviation from ideal spinel value of $125^{0}$.

\begin{figure*}[htbp!] 
%\centering    
%\includegraphics[width=1.0\textwidth]{fig2.pdf}
\includegraphics[width=0.8\textwidth]{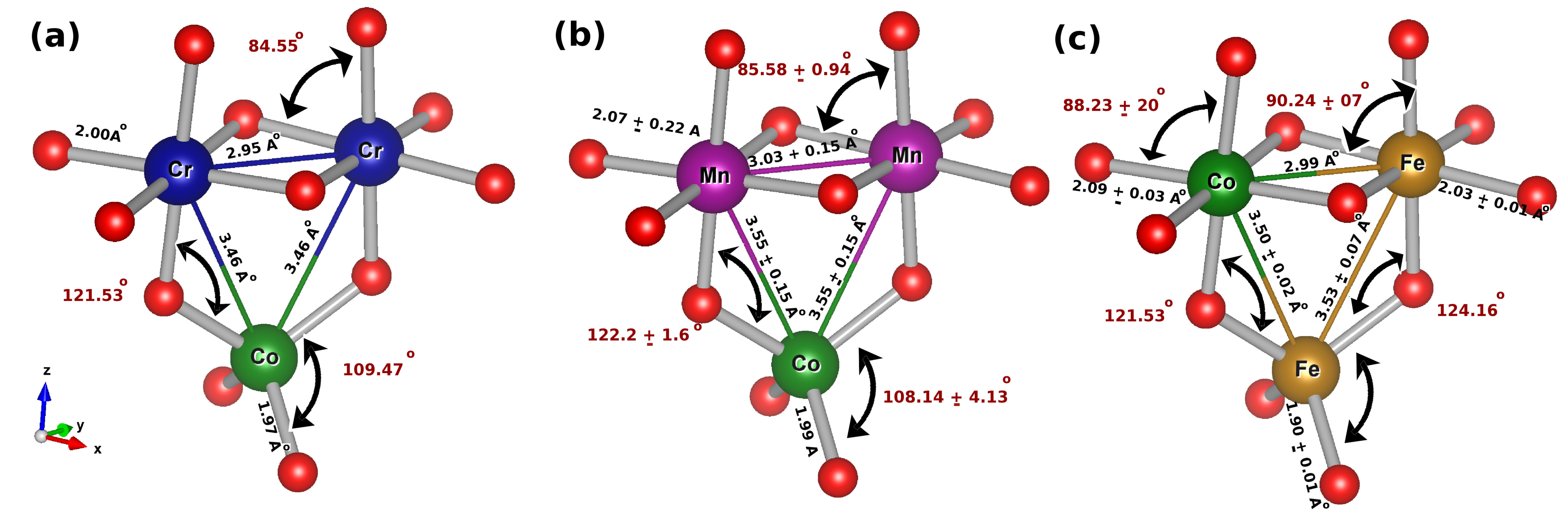}
\caption{The neighborhood around the tetrahedral and the octahedral sites
in (a) CoCr$_{2}$O$_{4}$, (b)CoMn$_{2}$O$_{4}$ and (c)CoFe$_{2}$O$_{4}$
along with the bond lengths and bond angles.}
\label{distorsion}
\end{figure*}

\begin{table*}[!htbp]
    \caption{\label{struc-tab} The lattice parameters of $CoB_2O_4$. $a,b,c$ are the lattice constants.
$x,y,z$ are the Oxygen parameters.} 
\begin{center}
        \begin{tabular}{|l|l|l|l|} \hline
      Structural         & $CoCr_2O_4$       &$CoMn_2O_4$ &$CoFe_2O_4$\\ 
      parameter          &                   &                  &          \\ \hline
          a(\AA)         & 8.35(8.34$^a$)    &  8.17(8.1$^{b,c}$)  &8.41(8.366$^d$) \\ \hline
          b(\AA)         & 8.35(8.34$^a$)    &  8.17(8.1$^{b,c}$)  &8.41(8.366$^d$) \\ \hline
          c(\AA)         & 8.35(8.34$^a$)    &  9.39(9.13$^b$,9.3$^c$)  &8.45(8.366$^d$) \\ \hline
          x              & 0.262(0.264$^a$)  & 0.255(0.23$^b$)           & 0.255(0.256$^d$) \\ \hline
          y              & 0.262(0.264$^a$)  & 0.255(0.23$^b$)            & 0.255(0.256$^d$)       \\ \hline
          z              & 0.262(0.264$^a$)  & 0.267(0.256$^b$)            & 0.255(0.256$^d$)       \\ \hline
            \end{tabular} 
        \end{center}
$a$: Reference \cite{CoCr2O4-exp}
$b$: Reference \cite{CoMn2O4-mag}
$c$: Reference \cite{comn2o4-struc}
$d$: Reference \cite{CoFe2O4-mag}
\end{table*}

\begin{figure}[htbp!] 
%\centering    
%\includegraphics[width=0.5\textwidth]{fig3.pdf}
\includegraphics[width=0.4\textwidth]{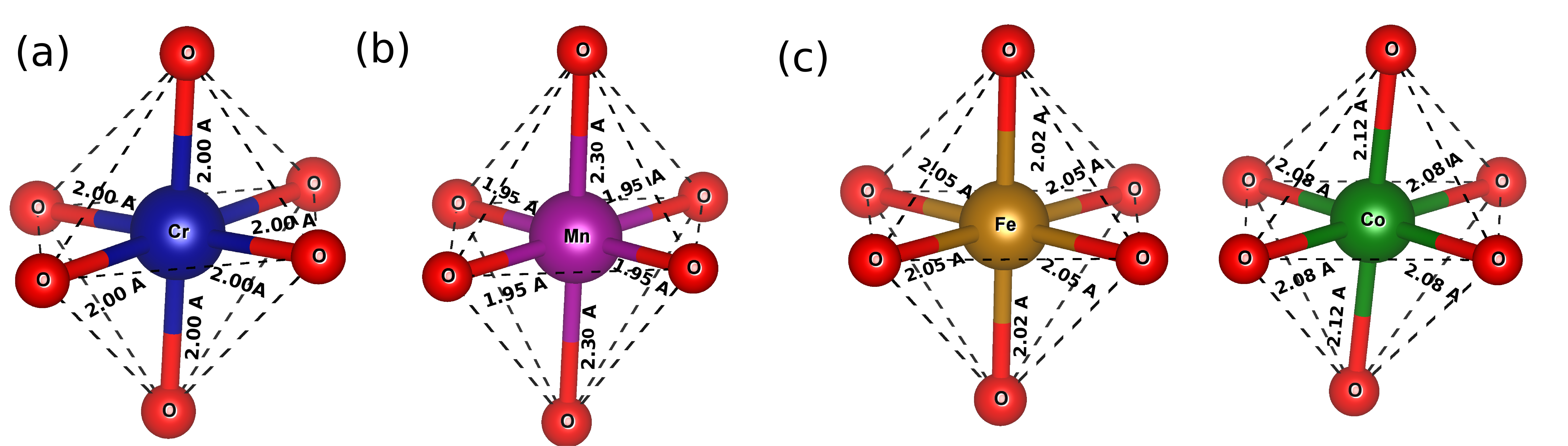}
\caption[Minion]{ The neighbourhood around the octahedral site of (a)$CoCr_2O_4$,(b) $CoMn_2O_4$ and (c)$CoFe_2O_4$ providing a close up view of the distortions 
associated with the octahedra.}
\label{distorsion}
\end{figure}
\begin{figure}[htbp!] 
\centering    
\includegraphics[width=0.5\textwidth]{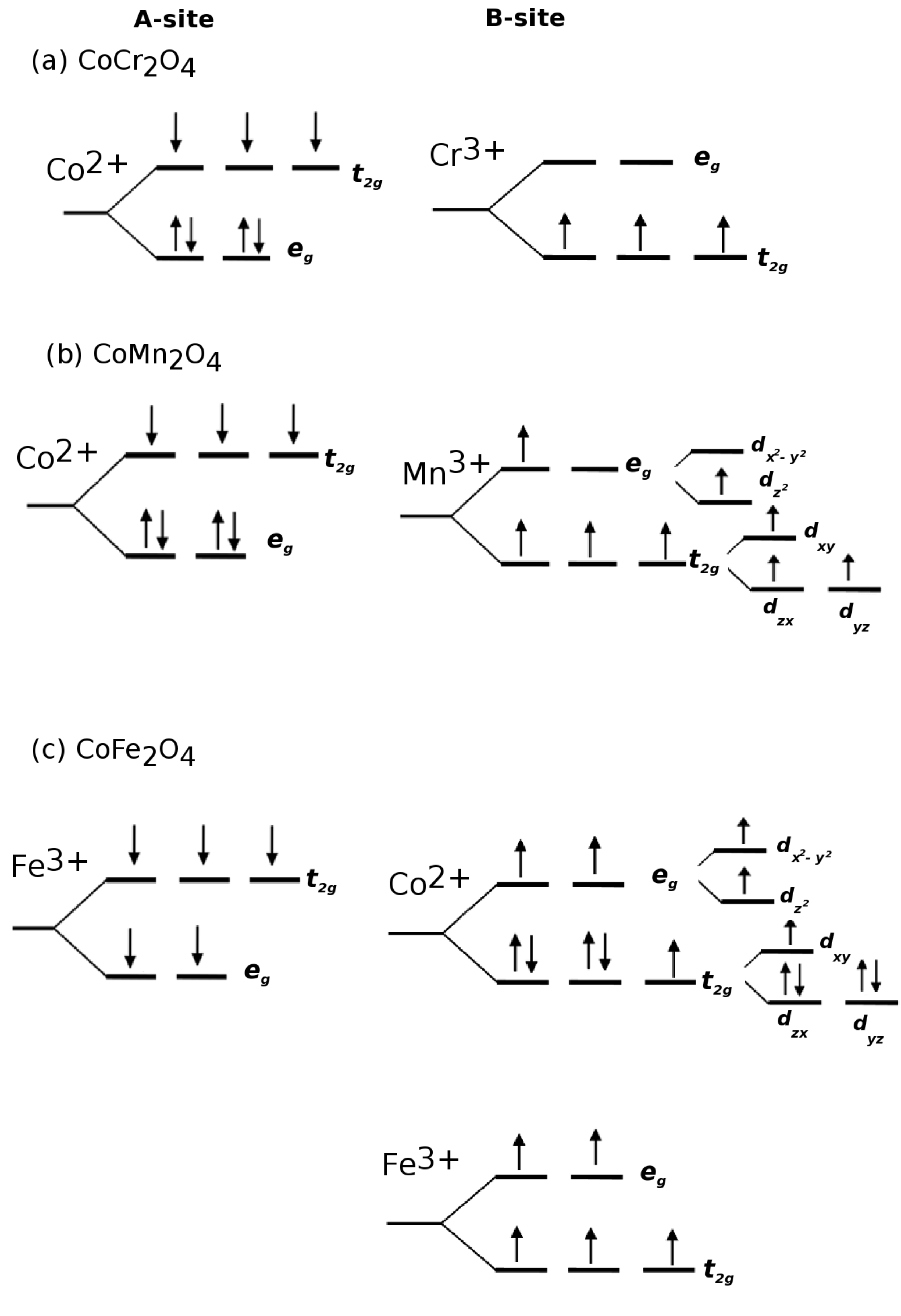}
\caption{\label{inversion}Electronic configurations for cations occupying A and B sites in CoB$_{2}$O$_{4}$ compounds}
\end{figure}
The energy level diagrams based upon the electronic configurations of the
cations occupying the tetrahedral and octahedral sites can explain the
observed trends in the structural aspects of the CoB$_{2}$O$_{4}$ compounds
considered here. Figure \ref{inversion} shows that in case of CoCr$_{2}$O$_{4}$
, the higher lying $e_{g}$ states of octahedral Cr$^{3+}$ are empty while
the $e_{g}^{1}$ configuration of Mn$^{3+}$ in CoMn$_{2}$O$_{4}$ implies
degeneracy associated with this level. In order to lift the degeneracy, the
symmetry of the crystal structure is lowered, giving rise to the tetragonal
ground state. The largeness in the tetragonal distortion can be understood 
from the $\left(t_{2g} \right)^{3}\left(e_{g} \right)$ configuration of the
Mn$^{3+}$ ions as explained by Dunitz and Orgel \cite{dunitz}. In case of
CoFe$_{2}$O$_{4}$, the degeneracy associated with the $t_{2g}$ orbitals of
the Co$^{2+}$ ions leads to the lowering of the cubic symmetry. The small
value of the distortion can once again be understood from the $\left(t_{2g}
\right)^{5}\left(e_{g} \right)^{2}$ of the Co atom in the octahedral site
\cite{dunitz}. 
 
\subsection{Electronic Structure and Magnetic moments}
\begin{figure*}[ht]
\includegraphics[width=15cm,  keepaspectratio]{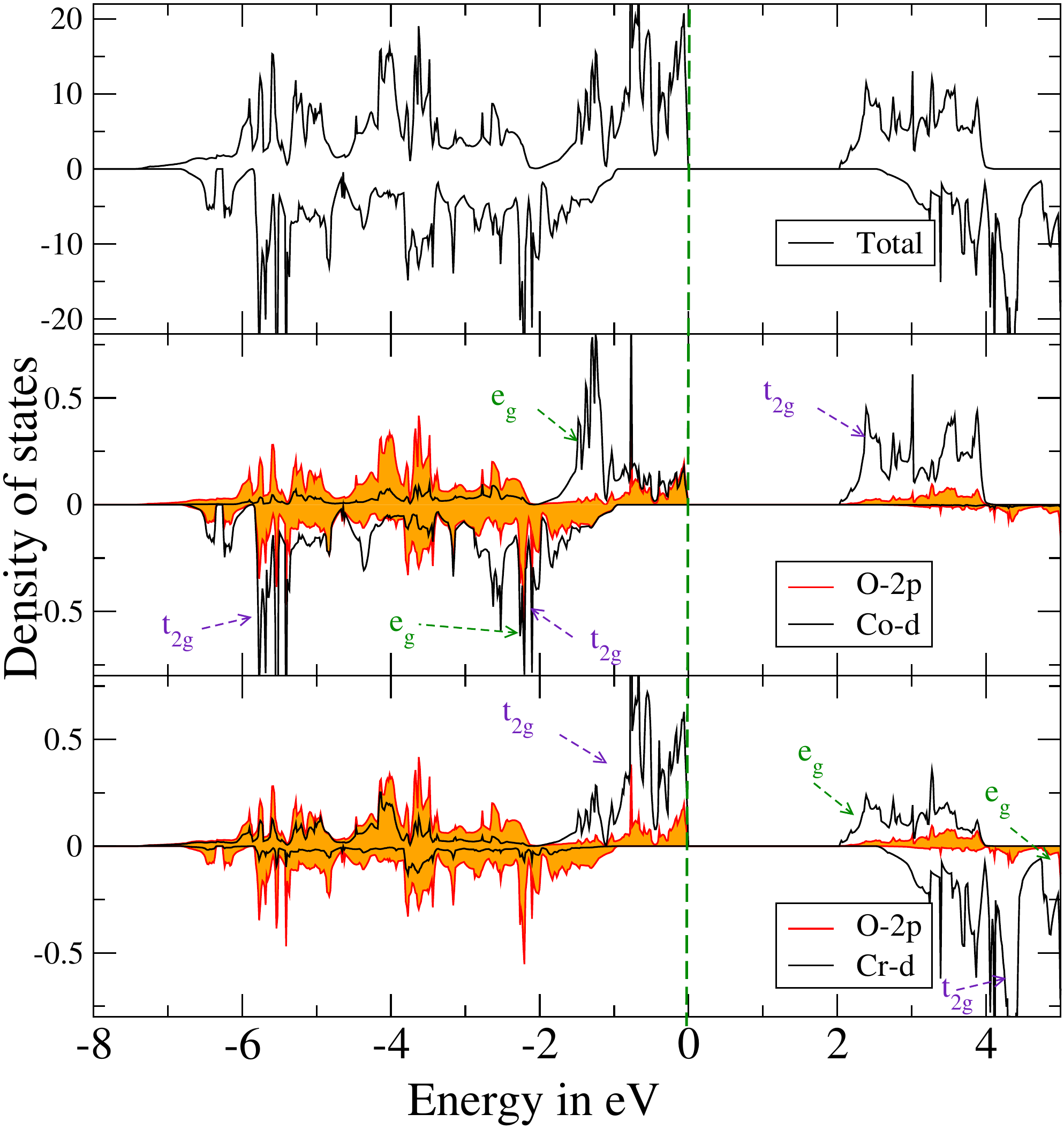}
\caption{\label{A-dos}The total(top panel) and site projected densities of states(A site in the middle panel and B site in the bottom panel, along with the
the anion contributions) 
for $CoCr_2O_4$. The energies are plotted with respect to Fermi level.}
\end{figure*}
The configurations of $d$ electrons at different cation sites play the most
important role in determining the electronic and magnetic properties of
spinel oxides.According to crystal field theory \cite{cf}, the electronic
configuration depends on the relative strengths of the crystal field(CF)
and intra-atomic exchange field(EX). In this subsection, we attempt an
understanding of the relation between the structural distortions and 
electronic structures in CoB$_{2}$O$_{4}$ compounds by investigating the
relative strengths of crystal field splitting and exchange splitting through
an analysis of the densities of states. Consequently,
this would lead to the understanding 
of the electronic and magnetic properties of these systems.
\begin{figure*}[ht]
\includegraphics[width=15cm,  keepaspectratio]{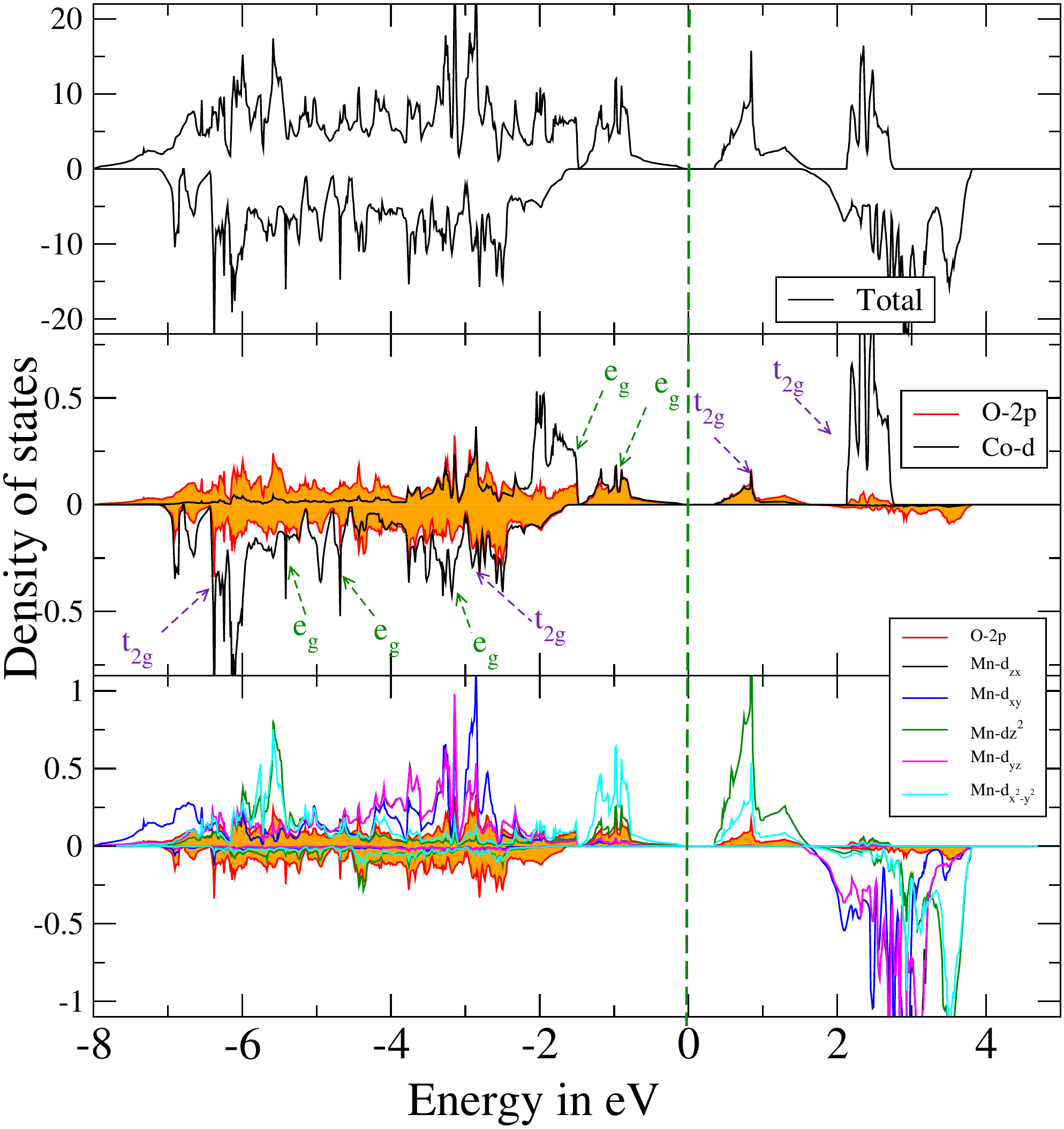}
\caption{\label{B-dos}The total(top panel) and site projected densities of states(A in the middle and B in the bottom panel along with anion contributions) 
for $CoMn_2O_4$. The Energies are plotted with respect to Fermi level.}
\end{figure*}

The schematic representation of electrons in $d$ levels in Fig.
\ref{inversion}, based upon crystal field theory, shows that in a 
tetrahedral crystal field, the $e_{g}$ levels lie lower than the $t_{2g}$
levels due to direct electrostatic repulsion between the $t_{2g}$
orbitals and the surrounding anion orbitals, while in an octahedral crystal
field, the order is reversed as the $e_{g}$ orbitals are repelled in this
case. Upon tetragonal distortion, the $e_{g}$ levels associated with the
octahedral sites further split into two levels with $d_{x^{2}-y^{2}}$
orbital at a energy higher than $d_{z^{2}}$, and the $t_{2g}$ level splits
into a higher $d_{xy}$ level and a lower doubly degenerate $d_{zy}(d_{zx})$
level when $c/a>1$. The spacings of these energy levels depend on the 
strengths of the crystal fields and the exchange fields.

\begin{figure*}[ht]
\includegraphics[width=15cm,  keepaspectratio]{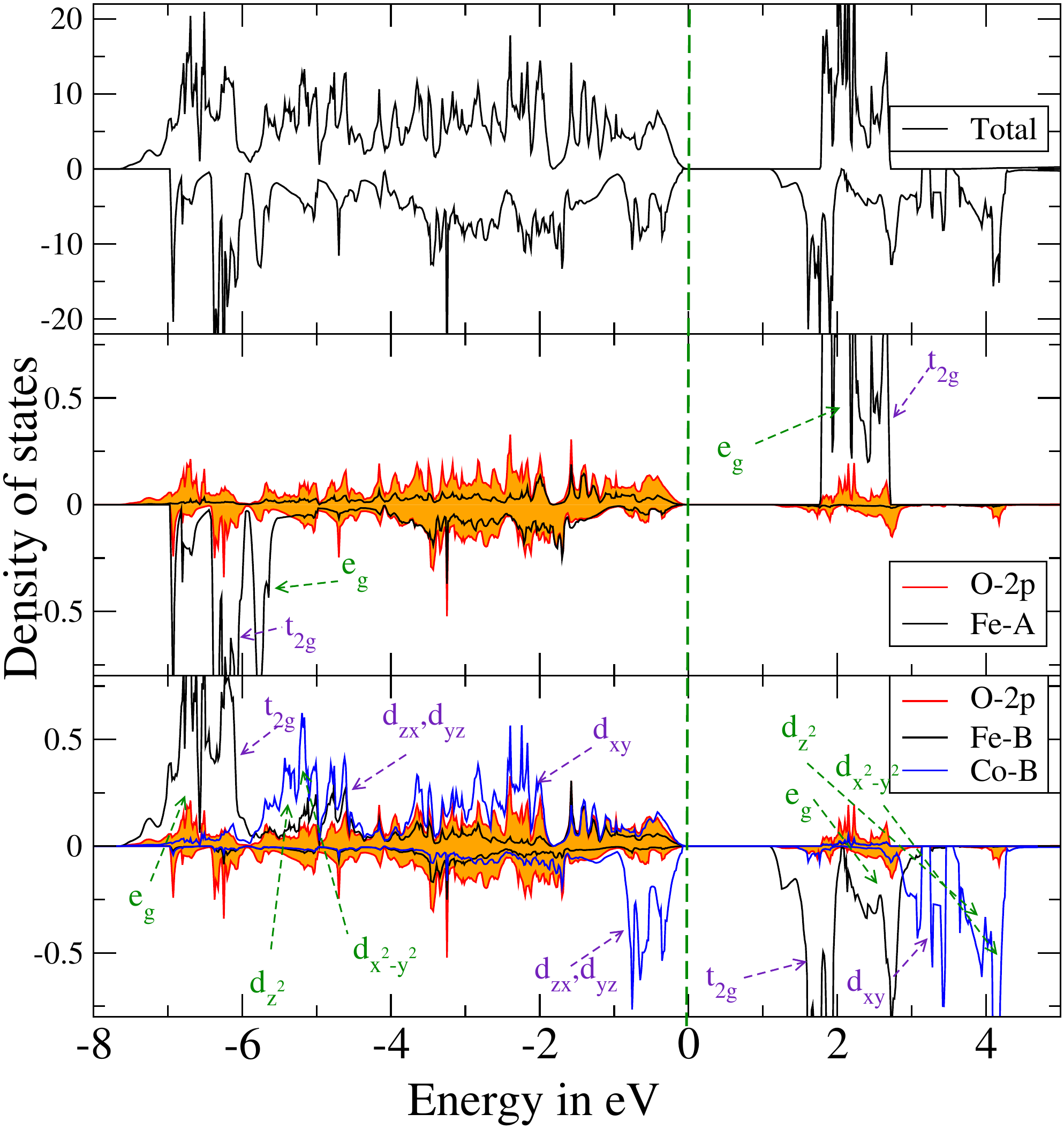}
\caption{\label{C-dos}The total(top panel) and site projected densities of states(A in the middle panel and B in the bottom panel along with anion 
contributions)
for $CoFe_{2}O_{4}$. The energies are plotted with respect to Fermi level.}
\end{figure*}
In Table \ref{table4} we present results for crystal field splitting
$(\Delta_{CF})$ and the exchange splitting $(\Delta_{EX})$ for the three
compounds. These results are obtained without incorporation of correlations
that is by setting $U_{eff}=0$. A comparison of results for the three 
\begin{table*}
    \caption{\label{table4} Exchange Splitting ($\Delta_{EX}$) and crystal field splitting ($\Delta_{CF}$) results for 
the three compounds.}  
\begin{center}
        \begin{tabular}{|c|c|c|c|c|c|c|c|c|c|c|c|c|} \hline
               System &  ion  &\multicolumn{4}{|c|}{ At tetrahedral site}&ion &  \multicolumn{4}{|c|}{At Octahedral site} \\ \hline
                      &       &$\Delta_{EX}^{e_g}$ &$\Delta_{EX}^{t_{2g}}$     & $\Delta_{CF}^{\downarrow}$ &$\Delta_{CF}^{\uparrow}$&&$\Delta_{EX}^{e_g}$ &$\Delta_{EX}^{t_{2g}}$  & $\Delta_{CF}^{\downarrow}$&$\Delta_{CF}^{\uparrow}$ \\ \hline
      $CoCr_2O_4$     & $Co^{2+}$   &  1.6  & 2.9   & 0.0  & 1.3  & $Cr^{3+}$ & 1.3 & 3.2 & 1.3 & 3.3 \\ \hline
     $CoMn_2O_4$& $Co^{2+}$   &  1.6  & 2.5   & 0.5  & 1.3  & $Mn^{3+}$ & 1.8  & 3.7 & 0.3 & 2.2  \\ \hline
 $CoFe_2O_4$& $Fe^{3+}$& 3.3& 3.1   & 1.0  & 0.8 & $Fe^{3+}$ & 3.5  & 3.7 & 1.7 & 1.9   \\   
               &      &       &       &    &              & $Co^{2+}$ & 3.1  & 2.4 & 1.9 & 1.2     \\ \hline 
            \end{tabular}
        \end{center}
\end{table*}
show that the stronger exchange splitting associated with the $t_{2g}$ states,
in comparison to crystal field splitting, is a common feature for the
compounds. This should lead to high spin states for both magnetic cations.
The results suggest that for all three materials, the five-fold $3d$ crystal
degeneracy is barely broken for the down spin($\downarrow$)channel of A site.
Therefore, irrespective of the chemical identity of the species at A site, this
spin band is nearly full. For CoCr$_{2}$O$_{4}$ and CoMn$_{2}$O$_{4}$, the
other $3d$ electrons mostly occupy the exchange split ($\Delta_{EX}^{e_{g}}=
1.6$ eV)$e_{g}$ up spin ($\uparrow$) channel. In case of CoFe$_{2}$O$_{4}$,
the exchange splitting of $e_{g}$ levels is double that of the other two
compounds. This is due to the fact that in CoFe$_{2}$O$_{4}$, the Fe$^{3+}$
occupies the tetrahedral sites having two electrons less than Co$^{2+}$
which occupies tetrahedral sites in the other two compounds, making the 
$e_{g}^{\uparrow}$ states nearly empty.The $t_{2g}^{\uparrow}$ states in 
all cases
are energetically higher and thus remain mostly unoccupied (Not shown here),
the exchange splitting associated with $t_{2g}$ states ($\Delta_{EX}^{t_{2g}}$)
being nearly constant, therefore.

The large exchange splitting of the $t_{2g}$ states at the B site for
CoCr$_{2}$O$_{4}$ and CoMn$_{2}$O$_{4}$ indicate that one of the spin bands is
nearly full and the other nearly empty. This indeed is the case with the
$t_{2g}^{\uparrow}$ bands nearly full in both cases(Not shown here). For
CoFe$_{2}$O$_{4}$, there are two different values of $\Delta_{EX}^{t_{2g}}$
corresponding to Fe$^{2+}$ and Co$^{2+}$. While the exchange splitting for Fe
is nearly same as that of Cr and Mn, it is smaller for Co. The reason is that
Co$^{2+}$ has two more electrons than Cr(Mn,Fe) and hence part of the down
($\downarrow$)spin channel is also occupied. The exchange splitting of $e_{g}$
states at the B site for CoCr$_{2}$O$_{4}$ and CoMn$_{2}$O$_{4}$ are much 
smaller as the $e_{g}$ band is nearly empty for both spins. One extra electron
in the $e_{g}$ band of CoMn$_{2}$O$_{4}$ in comparison to CoCr$_{2}$O$_{4}$
increases $\Delta_{EX}^{e_{g}}$ for the former. In case of CoFe$_{2}$O$_{4}$,
$\Delta_{EX}^{e_{g}}$ are comparable for both Co and Fe at the B site. This is
expected as both Co$^{2+}$ and Fe$^{3+}$  have one spin channel completely
full and the other completely empty.The extra electron in Mn$^{3+}$ $e_{g}$
band (as compared to Cr$^{3+}$) explains the reason behind smaller values of
crystal field splitting(for both bands) in CoMn$_{2}$O$_{4}$ than in
CoCr$_{2}$O$_{4}$. In CoFe$_{2}$O$_{4}$, $\Delta_{CF}^{\downarrow}$ for both
Co$^{2+}$ and Fe$^{3+}$ are the largest while $\Delta_{CF}^{\uparrow}$ are
smaller than that for Cr$^{3+}$ and Mn$^{3+}$ in the other two compounds.
The smaller values of $\Delta_{CF}^{\uparrow}$ are due to the fact that for both
cations, up($\uparrow$)bands are nearly full and therefore, the centers of
$t_{2g}^{\uparrow}$ and $e_{g}^{\uparrow}$ bands lie much closer than those of
Cr$^{3+}$ and Mn$^{3+}$ in the other two compounds. The largest crystal field
splittings for up($\uparrow$) bands of Co$^{2+}$ and Fe$^{3+}$ in 
CoFe$_{2}$O$_{4}$ are due to the facts that unlike Cr$^{3+}$ and Mn$^{3+}$ in
other two compounds, the $t_{2g}^{\downarrow}$ are not completely empty. Thus,
the separation between $t_{2g}^{\downarrow}$ and $e_{g}^{\downarrow}$ are
larger compared to the other two.
 \begin{table}[ht]
    \caption{\label{table2} The magnetic moments of $A(\mu_{A}),B(\mu_{B})$ cations and the total moment($\mu_{T}$) per formula
 unit in Bohr Magneton}
  \begin{center}
        \begin{tabular}{|c|c|c|c|c|c|} \hline
                System & A site & $\mu_A$  & B site      &  $\mu_{B}$         & $\mu_{T}$  \\ \hline
       $CoCr_2O_4$  & Co   &   -2.66   &Cr      &    2.94            &  2.95                 \\ \hline
       $MnCr_2O_4$&  Co & -2.68     &Mn    &    3.81            &  4.84                 \\ \hline
       $CoFe_2O_4$ & $Fe_{T}$    &  -3.98   & $Fe_{O}$  & 4.10 &  2.98                 \\ 
       & & & $Co_{O}$ & 2.66 &  \\ \hline
                      
            \end{tabular}
        \end{center}
\end{table} 
The outcome of the competition between the crystal field splitting
and the exchange splitting would affect the semiconducting band gap in
these materials. In case of CoCr$_{2}$O$_{4}$, equally strong exchange
splitting and crystal field splitting at Cr site decides the gap. For the
other two compounds, weaker crystal field splitting compared to the exchange
splitting at B sites should result in decrease in the band gap as the B
occupation changes from Cr to Mn and Fe. However, with the GGA only
calculations, ground states of CoMn$_{2}$O$_{4}$ and CoFe$_{2}$O$_{4}$ are
metallic.This is because of the presence of finite densities of $t_{2g}$
states at the Fermi level of B site atoms (Not shown here). 
This can be correlated to the
fact that without the inclusion of strong correlations, CoMn$_{2}$O$_{4}$
crystallizes in cubic structure, thus unable to obtain the symmetry breaking
of the $d$ orbitals and distribution of states on both sides of the Fermi
level. In case of CoFe$_{2}$O$_{4}$,the distortion at the Co site is small
and thus the local symmetry is barely broken, resulting in the localisation
of states at the Fermi level. The inclusion of electron-electron correlation
through GGA+U formalism produces the correct semiconducting 
ground states by introducing appropriate localisation of states, and
the correct crystal structures by breaking the degeneracies in the $e_{g}$
orbitals of Mn cations in CoMn$_{2}$O$_{4}$.The total and atom projected
densities of states calculated with GGA+U are presented in Figures \ref{A-dos},
\ref{B-dos} and \ref{C-dos}. In CoCr$_{2}$O$_{4}$,Co $t_{2g}^{\downarrow}$
electrons are localised around -6 eV and $e_{g}^{\downarrow}$ states are
localised around -2 eV. In the down spin ($\downarrow$) band, it is the $e_{g}$
states that primarily hybridise with the oxygen $p$ states. The $t_{2g}
^{\uparrow}$ Co states are unoccupied and centred around 2-4 eV, while the
2 $e_{g}$ electrons are localised between -1 eV and -2 eV in the up($\uparrow$) spin channel. This gives rise to a magnetic moment of about 3 $\mu_{B}$ at the
Co site (Table \ref{table2}). At the Cr site, the incorporation of strong
correlation does not have a very significant effect on the $t_{2g}$ states, 
the $t_{2g}^{\uparrow}$ states are still fully occupied with the electrons
localised near the Fermi level while the fully unoccupied $t_{2g}^{\downarrow}$ 
 states are pushed towards higher energies. The $t_{2g}$ states thus weakly
hybridise with the oxygen states. The $e_{g}$ band is nearly empty as it was
when calculations were done without incorporating $U_{eff}$. This results in
a magnetic moment of nearly 3$\mu_{B}$ for Cr atoms as well, with the sign
of the moment opposite to that of Co atoms (Table \ref{table2}).

In CoMn$_{2}$O$_{4}$(Fig. \ref{B-dos}), the $t_{2g}$ states of $Co$ are 
qualitatively similar to those in CoCr$_{2}$O$_{4}$, except that they are
shifted as a whole towards lower energy. The $e_{g}$ states at Co site
are more delocalised in comparison to those in CoCr$_{2}$O$_{4}$ leading
to more hybridisation between Co $e_{g}^{\downarrow}$ and oxygen $p$ states.
This, however, still gives rise to a magnetic moment of about 3$\mu_{B}$
at Co site, like CoCr$_{2}$O$_{4}$. The densities of states at the Mn site
is more interesting as the degeneracy of the $d$ states are lifted due to
large tetragonal distortion. According to crystal field theory, the higher
energy $e_{g}^{\uparrow}$ states should split into $d_{z^{2}}$ and $d_{x^{2}-
y^{2}}$ states with $d_{z^{2}}$ being occupied and $d_{x^{2}-y^{2}}$ 
participating in covalent bond formation with oxygen \cite{wickham}. In
Figure \ref{B-dos}(bottom panel), we indeed see this happening. The half-filled
$d_{z^{2}}^{\uparrow}$ gives rise to distinct peaks at around -6 eV and
at around 1 eV. The $d_{x^{2}-y^{2}}$ states are more delocalised, having
participated in hybridisations with oxygen $p$ states. The down ($\downarrow$)
spin channel is nearly empty, leading to a Mn moment of nearly 4 $\mu_{B}$
(Table \ref{table2}). The lifting of degeneracy has less effect on the $t_{2g}$
states except that they are more delocalised in comparison to the case 
when effect of electron electron correlation was absent and the crystal
structure was cubic. Overall, the extra electron in the $e_{g}$ band of Mn
(in comparison to Cr) gives rise to states closer to the Fermi level and thus
reduces the electronic band gap in CoMn$_{2}$O$_{4}$ in comparison to the
band gap in CoCr$_{2}$O$_{4}$ as expected. The calculated band gap in 
CoCr$_{2}$O$_{4}$ is 2.1 eV while that of CoMn$_{2}$O$_{4}$ is 0.33 eV.

In CoFe$_{2}$O$_{4}$, different sub-lattice occupancy due to the 'Inverted'
structure introduces qualitative differences of both A and B sites densities
of states (Fig. \ref{C-dos}). The introduction of strong correlations pushes
the unoccupied states in the spin up band towards higher energy. The 
completely filled spin down band has electrons extremely localised around
-6 eV to -7 eV allowing little hybridisations with the oxygen states.This
leads to a moment of 4 $\mu_{B}$ for Fe$_{T}$ atoms.
At the octahedral site, there are two types of atoms, Fe$_{O}$
and Co. We see qualitatively very different features of the states associated
with these two atoms. The $e_{g}^{\downarrow}$ band is completely empty for
either atoms. For Fe$_{O}$, the $t_{2g}^{\uparrow}$ band is completely full
and is extremely localised around -6 eV to -7 eV. Thus, irrespective of the
crystal environment, Fe states are localised in CoFe$_{2}$O$_{4}$. The Co 
states in the octahedral sites are, however, delocalised, particularly in
the spin up band. The distortions associated with the Co site, breaks the
$t_{2g}$ degeneracy only slightly. In the down spin band, the $d_{zx},d_{yx}$
states give rise to states near Fermi level(peaks close to -1 eV). There
were no such states associated with the octahedral sites in the other two
compounds. The extra electrons in Co, in comparison to Cr and Mn, are 
responsible for these states. This, in turn, reduces the band gap,
in comparison to CoCr$_{2}$O$_{4}$; the calculated band gap of 
CoFe$_{2}$O$_{4}$ being 1.14 eV. The $t_{2g}^{\uparrow}$ states of Co are 
delocalised, hybridising with the oxygen $p$ states.The magnetic moments
are, therefore about 3 $\mu_{B}$ and 4 $\mu_{B}$ for Co and Fe$_{O}$ in
CoFe$_{2}$O$_{4}$ (Table \ref{table2}). 

The comparative study of the electronic structure, thus, shows that the Co
states are delocalised irrespective of the crystal environment while the Fe
states are extremely localised in CoB$_{2}$O$_{4}$ compounds considered. This
would have important consequences on the magnetic exchange interactions and
therefore the spin structures of the pristine compounds as well as in doped
systems which have been investigated experimentally only recently \cite{
padam-apl,pss13,jap15}. In the next subsection we compute and discuss the
magnetic exchange interactions of these compounds.
\subsection{Magnetic Exchange Interactions} 
 \begin{table*}[ht]
    \caption{\label{table3} The magnetic exchange parameters ($J_{ij}$ in meV)
 and the Ferrimagnetic transition temperatures($T_{c}$ in K)of the three
compounds}
  \begin{center}
        \begin{tabular}{|c|c|c|c|c|c|c|c|} \hline
                System    &Type of & $J_{AA}$ &Type of &  $J_{BB}$ &Type of      &   $J_{AB}$             & $T_c$ \\
& AA pair & & BB pair & & AB pair & &  \\ \hline 
               $CoCr_2O_4$&Co-Co  &-0.56   &Cr-Cr  & -3.01 & Co-Cr &  -3.26       &  144(97$^a$)   \\  \hline               
               $CoMn_2O_4$& Co-Co &-0.29   &Mn-Mn  & -1.05(out of plane) & Co-Mn & -3.53  &  153(85$^b$) \\
& & & & -9.46(in plane)& & &      \\ \hline 
 $CoFe_2O_4$& $Fe_{O}-Fe_{O}$ &-2.06 & $Co_O-Co_O$ & 0.08 & $Fe_T-Co_O$ & -10.43        &        \\  
                         & & &  $Fe_O-Fe_O$ & -4.77 & $Fe_T-Fe_O$ & -21.65      & 1079(860$^c$)      \\ 
                         &  &  &  $Fe_O-Co_O$ &0.84  &  & &          \\  \hline
            \end{tabular}
        \end{center}
$a$: Reference \cite{CoCr2O4-exp}
$b$: Reference \cite{comn2o4-struc}
$c$: Reference \cite{CoFe2O4-mag}

        \end{table*} 

The magnetic exchange interactions are computed by mapping the GGA+U total
energies of different collinear spin configurations on to a Heisenberg
Hamiltonian \cite{dasjpd}. The results for the nearest neighbour
interactions are presented in Table \ref{table3}.The higher neighbour exchange
interactions are smaller by an order of magnitude and hence they are not
considered for discussions. The results show that the A-A interactions are
the weakest and therefore, is not expected to play any significant role. This
is consistent with the discussion by Kaplan \cite{kaplan}. In CoCr$_{2}$O$_{4}$
, the A-B and B-B exchange interactions are comparable. In this case B site
$t_{2g}$ orbitals are half-filled. So direct B-B interactions is possible,
apart from the super-exchange via oxygen atoms. Moreover, 
nearly empty $e_{g}$ orbitals
reduces the anion shielding as partial covalency via $e_{g}$ orbitals are
possible. However, since the A site $t_{2g}$ orbitals are half-filled and
the B site $e_{g}$ orbitals are nearly empty as seen from the densities
of states, the A-B interactions ceases to be the strongest.
Therefore, the A-B and B-B interactions are comparable for
CoCr$_{2}$O$_{4}$. Due to these two interactions being comparable, the spin
structure of CoCr$_{2}$O$_{4}$ is not collinear \cite{yafet} as has been
observed in the experiments. The calculation of the so called LKDM parameter
\cite{lkdm} which characterises the magnetic structure of cubic spinels
also confirmed this picture \cite{dasjpd}.
 In CoMn$_{2}$O$_{4}$, the A-site $t_{2g}$ orbitals
are half-filled while the B-site $e_{g}$ orbitals are less than half-filled
and degenerate. This should lead to a strong antiferromagnetic A-B interaction
which indeed is observed in our calculations ($J_{Co-Mn}=-3.48$ meV). As for
the B-B interaction, the B-site $t_{2g}$ orbitals are half-filled and so
direct B-B interactions, apart from super-exchange via oxygen is possible.
However, the strength of this interaction would depend on the inter-cation
distance \cite{wickham}. Due to the tetragonal distortion, the in-plane
Mn-Mn distances are short ($\sim 2.88 \AA$) and thus the direct interactions
are extremely strong$\sim -9.5$ meV (Table \ref{table3}). The out-of-plane
Mn-Mn distances are, on the other hand are $\sim 3.11 \AA$. This weakens the
B-B interactions and are only $\sim -1.05$ meV. This huge anisotropy in the
exchange interactions due to the tetragonal distortion leads to a complicated
non-collinear spin structure \cite{CoMn2O4-mag}. This anisotropy in the
CoMn$_{2}$O$_{4}$ was discussed earlier \cite{magbook} but was never 
calculated from first-principles. Our calculations provided the quantitative
credence to the original idea.

In case of CoFe$_{2}$O$_{4}$, the overwhelmingly dominant interactions are
the A-B interactions as is seen from our calculations. This explains the
reason behind a collinear spin structure of this system. The reason 
behind this strong antiferromagnetic A-B interaction is that the A site
$t_{2g}$ orbitals and B site $e_{g}$ orbitals (for both Fe and Co at B site)
are half-filled. The reason behind
a weak Co-Co and Co-Fe interactions at the B site is that the Co $t_{2g}$
orbitals are more than half-filled and degenerate and hence no direct B-B
interactions occurs between them. Fe$_{O}$-Fe$_{O}$ interactions are
stronger than these two as the $t_{2g}$ orbitals of Fe$_{O}$ are
half-filled and hence direct interactions are possible. However, since
the Fe$_{O}$ $e_{g}$ orbitals are also half-filled, partial covalency
with anions is also not possible(as seen in the extremely localised Fe
densities of states), leading to a large anion shielding
and a subsequent reduction in the B-B interaction in comparison to the
A-B interaction.

The ferrimagnetic transition temperatures ($T_{C}$) are calculated using these
exchange interactions and under mean field approximation, as done in
Ref. \cite{dasjpd}. Our calculated results and a comparison with experimental
values are presented in Table \ref{table3}. As expected, the critical
temperatures calculated by a Mean field approximation overestimates the
values. However, the calculated values reproduce the trends seen in the
experiments with CoFe$_{2}$O$_{4}$ having a large $T_{C}$ due to very large
values of $J_{AB}$ while the other two have rather small values of $T_{C}$
due to competing $J_{AB}$ and $J_{BB}$ leading to non-collinear magnetic
structures.  

\section{Conclusions}
We have performed a systematic investigations into the structural and
magnetic properties of CoB$_{2}$O$_{4}$ magnetic spinels by changing the
B cation using first-principles Density functional theory based methods.
The understanding of the properties of these compounds is done by quantifying
the relative strengths of the crystal field effect and the exchange effect
through an analysis of their electronic structures. We find that the
electron-electron interactions of the magnetic cations play a very important
role and without the incorporation of this, the correct ground state
structures cannot be obtained. The strong electron correlations are responsible
for significant local structural distortions at the octahedral site and
global tetragonal distortions for CoMn$_{2}$O$_{4}$. These are responsible
for the trends in the electronic properties such as the band gap. The
electronic structures of these compounds are significantly different as the
B site cation is changed. This, in turn, affects the inter-atomic magnetic
exchange interactions considerably and is responsible for very different
spin structures of these systems. In this work, for the first time, 
understanding of the trends in the magnetic properties are attempted through
proper quantification of the associated quantities and by providing
necessary explanations from the trends in the local structural parameters and
the electronic structures. The results hold immense significance with 
regard to the recent experimental results on Co(Cr$_{1-x}$Fe$_{x}$)$_{2}$O$_{4}$
and Co(Cr$_{1-x}$Mn$_{x}$)$_{2}$O$_{4}$ systems, where the chemical 
properties of the third magnetic atom in CoCr$_{2}$O$_{4}$, the sub-lattice
occupancies and the structural distortions are thought to be giving rise
to interesting functional properties. In future communications, these issues
will be addressed.
\section{Acknowledgments}
The computation facilities from C-DAC, Pune, India
and from Department of Physics, IIT Guwahati funded under the FIST programme of DST, India
are acknowledged.


\begin{thebibliography}{99}
\bibitem{spinel1} N.V.Kuleshov,V.P.Mikhialov,and V.G.Scherbitsky, 
{\it Proc. SPIE}, {\bf 175},2138 (1994)
\bibitem{spinel2} A.Jouini,A.Yoshikawa,A.Guyot,A.Breiner,T.Fukuda,
and G.Boulon, {\it Opt. Mater.} {\bf 30}, 47 (2007)
\bibitem{spinel3} K.G. Tshabalala,S.H.Cho,J.K.Park,S.S.Pitale,I.M.Nagpure,
R.E.Kroon,H.C.Swart,and O.M.Ntwaeaborwa {\it J. Alloys. Compds.} 
{\bf 509}, 10115 (2011)
\bibitem{spinel4} Y.X.Li,P.J.Niu,L.Hu,X.W.Xu,and C.C.Tang {\it J. Lumin.} 
{\bf 129}, 1204(2009)
\bibitem{spinel5} P.J.Deren, K.Maleszka-Baginska, P.Gluchowski, and M.A.Malecka 
{\it J.Alloys.Compds.} {\bf 525}, 39 (2012)
\bibitem{cocr2o4} Y.Yamasaki, S.Miyasaka, Y.Kaneko, J.-P.He, T.Arima and 
Y. Torakuma, {\it Phys. Rev. Lett.} {\bf 96}, 207204 (2006)
\bibitem{cocr2o4-1}K.Tomiyasu,J.Fukunaga and H.Suzuki, {\it Phys. Rev.} {\bf B
70}, 214434 (2004)
\bibitem{cocr2o4-2}L.J.Chang, D.J.Huang, W.-H Li,S.-W. Cheong, W. Ratcliff
and J.W.Lyn, {\it J.Phys.Condens. Matt.} {\bf 21}, 456008 (2009)
\bibitem{padam-apl}R.Padam,S.Pandya,S.Ravi,A.K.Nigam,S.Ramakrishnan,
A.K.Grover and D.Pal, {\it Appl. Phys. Lett.} {\bf 102}, 112412 (2013)
\bibitem{padam-aip}R.Padam,S.Pandya,S.Ravi,A.K.Nigam,S.Ramakrishnan,
A.K.Grover and D.Pal, {\it AIP Conference Proceedings} {\bf 1512}, 
1112 (2013)
\bibitem{padam-thesis}R.Padam, Ph.D.Thesis, IIT Guwahati (2014)
\bibitem{pss13}H.-Zhang, W.-Wang, En-Liu, X.-Tang, G.-Li, H.-Zhang and
G.-Wu, {\it Phys. Stat. Solidi} {\bf B 250}, 1287 (2013)
\bibitem{jap15}H.G.Zhang, Z.Wang, E.K.Liu, W.H.Wang, M.Yue and G.H.Wu,
{\it J. Appl. Phys.} {\bf 117}, 17B735 (2015)
\bibitem{dft}P. Hohenberg and W. Kohn, {\it Phys. Rev.} {\bf B136}, 864 (1964); W.Kohn and L.J.Sham, {\it Phys. Rev.} {\bf A140}, 1133(1965)
\bibitem{ederer}C.Ederer and M. Komelj, {\it Phys. Rev.} {\bf B76}, 064409(2006)
\bibitem{dasjpd}D.Das and S. Ghosh, {\it J. Phys. D:Appl. Phys.} {\bf 48},
425001 (2015)
\bibitem{ederer1}D.Fritsch and C.Ederer, {\it Phys. Rev.} {\bf B82}, 104117 (2010)
\bibitem{jphysd} Y.H.Hou, Y.J.Zhao, Z.W.Liu, H.Y.Yu, X.C.Zhong, W.Q.Qiu, D.C.Zeng and
L.S.Wen, {\it J.Phys.D: Appl. Phys.} {\bf 43}, 445003 (2010)
\bibitem{biplab} S.Ganguly,R.Chimata and B.Sanyal, {\it Phys. Rev.} {\bf B92},
224417 (2015)
\bibitem{dftu}V.I.Anisimov, F.Aryasetiawan and A.I. Liechtenstein, {\it J. Phys. Condens. Matt.} {\bf 9}, 767 (1997)
\bibitem{dudarev}S.L.Dudarev,G.A.Botton,S.Y.Savrasov,C.J.Humphreys and A.P.Sutton, {\it Phys. Rev.}{\bf B57}, 1505 (1998)
\bibitem{J}I.Solovyev,N.Hamada and K.Terakura, {\it Phys. Rev.}{\bf B53}, 7158 (1996)
\bibitem{paw}P.E.Blochl, {\it Phys. Rev.} {\bf B50}, 17953 (1994)
\bibitem{vasp} G.Kresse and J. Furthmuller, {\it Comput. Mater. Sci.} {\bf 6}, 15 (1996)
\bibitem{pbe}J. P. Perdew, K. Burke, and M. Ernzerhorf, {\it Phys. Rev. Lett.} {\bf 77}, 3865 (1996)
\bibitem{menyuk}N.Menyuk, K.Dwight and A.Wold, {\it J. Phys.(Paris)} {\bf 25}, 528 (1964)
\bibitem{CoMn2O4-mag}  B. Boucher, R. Buhl and M. Perrin , {\it J. Appl. Phys.} {\bf 39},632 (1968)
\bibitem{CoFe2O4-mag}  J. Teillet F. Bouree and R. Krishnan {\it J. Magn. Magn. Mater.} {\bf123},93-96 (1993)
\bibitem{CoCr2O4-mag} K. Tomiyasu, J. Fukunaga, and H. Suzuki, {\it Phys. Rev.} {\bf B70},214434 (2004)
\bibitem{CoFe2O4-moss}S.J.Kim,S.W.Lee and C.S.Kim, {\it Jpn. J. Appl. Phys.},
{\bf 40}, 4897 (2001)
\bibitem{CoCr2O4-exp}G. Lawes, B.Melot, K.Page, C.Ederer, M.A.Hayward, T.Proffen and R. Seshadri, {\it Phys. Rev.}{\bf B74}, 024413 (2006)
\bibitem{comn2o4-struc}D.G.Wickham and W.J.Croft, {\it J. Phys. Chem. Solids.}
{\bf 7}, 351 (1958)
\bibitem{dunitz} J.D.Dunitz and L.E.Orgel, {\it J. Phys. Chem. Solids} {\bf 3}, 20 (1957)
\bibitem{cf} I.B.Berusker, {\it Electronic structure and Properties of Transition Metal
Compounds: Introduction to the Theory} (New York: Wiley) (1996)
\bibitem{wickham}D.G.Wickham and J.B.Goodenough, {\it Phys. Rev.} {\bf 115},
1156 (1959)
\bibitem{kaplan}T.A.Kaplan, {\it Phys. Rev.}, {\bf 119}, 1460 (1960)
\bibitem{yafet}Y.Yafet and C.Kittel, {\it Phys. Rev.}, {\bf 87}, 290 (1952)
\bibitem{lkdm}D.H.Lyons, T.A.Kaplan, K.Dwight and N. Menyuk, {\it Phys. Rev.}{\bf 126}, 540 (1962)
\bibitem{magbook}J.B.Goodenough, {\it Magnetism and the Chemical Bond}
(John Wiley) (1963)
\end{thebibliography}
\end{document}